\documentclass[a4paper,11pt]{article}
\usepackage{amsfonts}
\usepackage{latexsym}
\usepackage{amsmath}
\usepackage{amssymb}
\usepackage{amssymb}
\usepackage{slashed}
\usepackage{upgreek}
\usepackage{mathrsfs}
\usepackage{bbm}

\usepackage{amsthm}

\theoremstyle{remark}

\newtheorem*{rmk*}{Remark}

\theoremstyle{remark}

\usepackage[utf8]{inputenc}
\usepackage[english]{babel}

\theoremstyle{definition}

\allowdisplaybreaks

\usepackage{relsize}
\usepackage{graphicx}
\usepackage{comment}

\renewcommand{\thefootnote}{\fnsymbol{footnote}}

\def\appendix#1{\addtocounter{section}{1}\setcounter{equation}{0}
\renewcommand{\thesection}{\Alph{section}}
\section*{Appendix \thesection\protect\indent \parbox[t]{11.15cm}{#1}}
\addcontentsline{toc}{section}{Appendix \thesection\ \ \ #1}}

\font\mybb=msbm10 at 11pt

\def\bb#1{\hbox{\mybb#1}}

\def\bZ {\bb{Z}}
\def\bR {\bb{R}}

\def\rsq {]\kern -2.0pt]}
\def\lsq {[\kern -2.0pt[}

\def\be{\begin{equation}}
\def\ee{\end{equation}}

\def\rsq {]\kern -2.0pt]}
\def\lsq {[\kern -2.0pt[}

\newcommand{\bea}{\begin{eqnarray}}
\newcommand{\eea}{\end{eqnarray}}

\usepackage[usenames,dvipsnames,svgnames,table]{xcolor}	
\usepackage[normalem]{ulem}				
\usepackage{microtype}
\usepackage[colorlinks, citecolor=blue, linkcolor=blue, urlcolor=Maroon, filecolor=Maroon, linktocpage=true]{hyperref} 

\usepackage{chngcntr}
\counterwithout{equation}{section}

\begin{document}

\begin{center}
\vspace*{-1.0cm}
\begin{flushright}
\end{flushright}

\vspace{2.0cm} {\Large \bf Brane flows } \\[.2cm]

\vskip 2cm
   Georgios  Papadopoulos${}^*$ and Kostas Skenderis${}^\dagger$
\\
\vskip .6cm

\begin{small}
${}^*$\textit{Department of Mathematics
\\
King's College London
\\
Strand
\\
 London WC2R 2LS, UK}\\
\texttt{george.papadopoulos@kcl.ac.uk}
\\*[.6cm]
${}^*$\textit{STAG Research Centre  \& Mathematical Sciences
\\
University of Southampton
\\
Southampton, SO17 1BJ, UK}\\
\texttt{K.Skenderis@soton.ac.uk}
\end{small}
\\*[.6cm]
\end{center}

\vskip 2.5 cm

\begin{abstract}
\noindent

Based on effective D-brane actions, we present a generalisation of the Ricci flow that includes the flow of a theory with a $n$-form field strength for $n\geq 0$. This is a generalisation of both the Ricci flows and the generalised Ricci flows.
Following Perelman, we show that flows that keep a suitable field-dependent volume fixed are monotonic. We also show that all steady brane flow solitons are gradient solitons and use this to demonstrate that on some occasions this implies the existence of a Killing vector field that leaves all the other fields invariant. Particular cases of gradient solitons are NS5 and D5 branes, and the volume which is kept fixed in these cases is the T-duality invariant volume (NS5 brane) or its S-dual (D5 brane).
We also generalise the above analysis to gravitational actions coupled to form gauge potentials  that  also exhibit a Chern-Simons type term. We find an alteration is required in the adaptation of Perelman's modification to this case, which yields a new functional that also exhibits a Chern-Simons term.  Under suitable assumptions, we proceed to prove the monotonicity of the flow and that all steady flow solitons are gradient solitons. We also explore the consequences of the last statement on the geometry of solitons.

\end{abstract}



\newpage

\renewcommand{\thefootnote}{\arabic{footnote}}



 \setcounter{section}{0}

\numberwithin{equation}{section}

\section{Introduction}\label{intro}

One of the most celebrated recent results in geometric topology is the Perelman's proof of the Poincar\'e conjecture \cite{Perelman}  based of the Hamilton's program of Ricci flows \cite{Hamilton}. This states that all compact simply connected 3-dimensional manifolds are diffeomorphic to the standard 3-sphere. It has been clear from the very beginning that these results must have applications in physics as the Ricci flow is the same as the 1-loop renormalisation group flow of (bosonic) 2-dimensional sigma models with a metric coupling \cite{Friedan}. This relation  has been rather fruitful. Using the renormalisation group flow of 2-dimensional sigma models with a $B$-field coupling,  it allows for the generalisation of the Ricci flows\footnote{In the mathematics literature, these are called generalised Ricci flows. As we shall propose another generalisation and keeping on with the brane theme of this paper, we shall refer to them here as fundamental string flows.} to include, apart from the flow of the metric, the flow a 2-form gauge potential.  The surprising fact is that many of the techniques applied by Perelman to control the behaviour of the  Ricci flow have a generalisation in this more general setting that includes the $B$-field \cite{OSW, OSW2}, see also \cite{Tseytlin2, Huhu} for related work. Amongst the results that have been obtained in physics are first the use of Perelman style techniques to give a geometric proof of Zamolodchikov's c-theorem \cite{Zam} at least in the context of sigma models \cite{OSW2}. This states that the renormalisation group flow of unitary  2-dimensional quantum field theories is monotonic. Another significant result is a geometric proof for sigma models \cite{gpew} of Polchinski's statement that all scale invariant, unitary, 2-dimensional field theories with a discrete spectrum of operator dimensions are conformally invariant \cite{Polchinski}. This generalisation to include a $B$-field has also some profound applications in geometry, especially in the investigation of manifolds with a special structure associated with a  connection with skew-symmetric torsion, see e.g. \cite{GFS, abls, gpheterotic, bfgv, gp1}.  Further generalisations\footnote{Although our work is related to these generalisations, our direction of investigation is different as we focus on the monotonicity of the flows and the relation between the steady and gradient solitons. Also, we use different functionals to establish our results.} of the Ricci flow have also been considered in the literature based on the consideration of $D=11$ and type II  supergravities, and the Einstein-Maxwell system. These have been used  to investigate the existence of solutions to these theories, to explore the swampland conjecture and to study the stability of black holes \cite{fei, Luest1, Luest2}.

In this paper, we shall explore another generalisation of the Ricci flow that apart from the flow of the metric $g$ of a manifold $M^D$ involves the flow of a $(n-1)$-form gauge potential $A$ on $M^D$ as well -- unless it is otherwise explicitly stated, $M^D$ has Euclidean signature and it is compact without boundary.  Such a flow will be investigated in conjunction with
 the D$p$-brane effective actions
 \be
S=- \int_{M^D} d^Dx\, \sqrt{g}\, e^{2a\Phi}\, \big(R+a^2 b (\partial\Phi)^2+ c F^2\big)~,
\label{dpbrane}
\ee
of \cite{Boonstra:1998mp, ksmt}, written $D$-dimensions, where $a, b, c$ are (coupling) constants that are related to $p$ \footnote{Note that the constants $a$ and $c$ may be removed by a rescaling of $\Phi$ and $A$, respectively. The (absolute value of the) overall constant in front of the action can also be adjusted arbitrarily by shifting $\Phi$ by a constant, and we used this freedom to set it to $-1$. The overall sign will not play a role for us; it is normally fixed such that the Euclidean action is positive for physical excitations, and this is the sign we choose in \eqref{dpbrane}.}, see also the earlier pioneering work in \cite{Duff:1990hb, Duff:1991pe,Duff:1994fg}. 
The objective is twofold. The first part is to identify the flows, which are monotonic. As in Perelman's discussion \cite{Perelman} one needs to modify the functional (\ref{dpbrane}) such that a suitable volume is kept fixed along the flow.
The second is to show that all steady solitons are gradient solitons -- in the case of sigma models mentioned above, this statement  is equivalent to proving that  all scale invariant models are actually conformally invariant.

We find that these two objectives hold for a variety of flows provided that $M^D$ is compact without boundary, the flows exist at least for some finite time $t$ and the fields are smooth. The first possibility is a straightforward generalisation of the fundamental string flow that instead of the $B$ field one allows the flow of a $(n-1)$-form gauge potential and considers $\Phi$ as the gradient scalar, see (\ref{flow1}) below. We find that under these assumptions both objectives are met provided that the constant $b=4$ in (\ref{dpbrane}).  It turns out that for this value of $b$, $p=5$ and (\ref{dpbrane}) is the effective action of D5-branes. As D5-branes are dual to NS5-branes and the latter are solutions of the fundamental string flow, it is not surprising that such a restriction arises as this can be interpreted as an indication that such flows are related  to the fundamental string flows. Indeed, this is the case provided that $F$ is a 3-form field strength. But in general, there is no restriction on the degree of $F$.  Therefore, the flow equations (\ref{flow1}) generalise the Ricci flow as well as the fundamental string flow. It is also significant to mention that the above analysis holds provided that the coupling $c F^2$ is replaced with $\sum_k c_k F_k^2$, where $F_k$ is a $k$-form field strength, $d F_k=0$.  This will be proved useful in the investigation of more general flows.
We also point out that the relation between steady and gradient solitons has geometric consequences. In particular, we demonstrate that some of these solitons must exhibit a Killing isometry that leaves the form field strength $F$ invariant.

Next, we consider more general flows that do not  require the restricts $b=4$. We take two approaches to this. In the first, we modify (\ref{dpbrane}) by adding  another scalar field $\Psi$. Then we demonstrate, after performing a suitable linear combination of $\Phi$ and $\Psi$, that we can identify a suitable gradient scalar such that the resulting flow is monotonic and all its steady solitons are gradient solitons.  It turns out that the new effective action is like in (\ref{dpbrane}) with the addition of an axion field, {\it i.e.} the addition of an 1-form field strength. It turns out that with the addition of the axion, it is no longer required that $b=4$.  However, apart from the flow equations of
$g$ and $A$ an additional flow equation must be added that of the axion.

Alternatively, one can start with specifying the gradient scalar in the flow and then write down an ansatz  for an  action that describes the coupling of the gradient scalar  to the rest of the fields. One of the novelties of this approach is that the adaptation of Perelman's argument involves the eigenvalues of a Strum-Liouville type of equation instead of the usual Schr\"odinger one, which luckily has the required properties for the existence of a suitable  ground state. It turns out that this second approach yields  the same result as the previous one.

Another generalisation  we present is flows based on an action $S_T$ that exhibits an additional  Chern-Simons type term, {\it i.e.},
\be
S_T= S+ S_{\mathrm{CS}}~,~~~S_{\mathrm{CS}}=e \int_{M^D} \, A\wedge^k F~,
\ee
where $S$ is given in (\ref{dpbrane}).
 The presence of this new term  can be motivated by the actions of supergravity theories, like for example that of 11-dimensional supergravity, that exhibit such couplings. However, here the consistency of the construction that follows requires  for the form $A\wedge^k F$ to be {\it globally defined} on $M^D$. This is for the action $S_{\mathrm{CS}}$ to be well-defined on $M^D$.  The addition of  $S_{\mathrm{CS}}$ modifies the flow of the gauge potential $A$ but otherwise leaves the flows of the other fields unchanged. Remarkably, a Perelman's argument is still possible for the new theory though the functional that it is defined after the modification  has a Chern-Simons term.  Furthermore, one can demonstrate monotonicity of the flows and that all steady solitons of these flows are gradient solitons.

The paper is organised as follows:  In section two, we give the the flow equations of the metric and that of a general $(n-1)$-form gauge potential. Then motivated by D-brane techniques, we describe actions for these fields  whose field equations are those that describe the gradient solitons of these flows. We also give the restrictions on the couplings such that the gradient soliton scalar is identified with the D-brane dilaton and demonstrate that the only allowed coupling is that of type II 5-branes.  Then, after a Perelman style modification, we demonstrate that the flows are monotonic and that all steady solitons of the flows are gradient solitons.  We also point out that the constraints required for the Perelman style modification on the spacetime volume are either T-duality invariant or related to the T-duality invariant volume by S-duality. In section three, we  introduce an axion in the theory and find that the restriction of the couplings to those of D5-branes in the action can be replaced with a weaker condition -- we approach the problem in two different ways leading to the same result at the end.  In section four, we generalise our construction to flows for which  their gradient solitons are described by the field equations of  actions that include  a Chern-Simons term. Such a construction is motivated by supergravity actions, like for example that of 11-dimensional supergravity. Surprisingly, we find that one can perform a Perelman style modification to these actions and the resulting functional can be used to prove the monotonicity of the flows. One can also prove that all steady solitons of these flows are gradient solitons.  Finally in appendix A, we justify the restriction we have to make for the Chern-Simons term to be globally defined in order to carry out the analysis in section four.

\section{Flows for branes}\label{Perelmanstyle}

\subsection{The flow, steady and gradient solitons}\label{ssec:one}

Let us consider  a theory with a metric $g$ and $n$-form field strength $F$, $dF=0$, $F=dA$ locally. The first set of the  flow equations with parameter $t$ that we shall begin to explore is
\begin{align}
\frac{d}{dt} g_{\mu\nu}&=-\big(R_{\mu\nu}+ n c F_{\mu\lambda_1\dots \lambda_{n-1}} F_{\nu}{}^{\lambda_1\dots \lambda_{n-1}}-2 a \nabla_\mu\partial_\nu \Phi\big)
\cr
\frac{d}{dt} A_{\lambda_1\dots \lambda_{n-1}}&= -2c \Big(\nabla^\mu F_{\mu\lambda_1\dots \lambda_{n-1}}+2 a \nabla^\mu \Phi F_{\mu\lambda_1\dots \lambda_{n-1}}\Big)
\label{flow1}
\end{align}
where  $\Phi$ is a function -- the flow gradient scalar. The constants $a$ and  $c$ that we have introduced in the flow equations above may not be related to those of the D$p$-brane action in (\ref{dpbrane}).  But in anticipation of using this functional to prove several properties of the flow, we have a priori made the identification.  The flows above are a generalisation of the Ricci flows as well as a generalisation of the  generalised Ricci flows.
We have not introduced a flow equation for $\Phi$.  It is possible to do so but we shall keep the form of the flow equations as in (\ref{flow1}). It will become apparent in the discussion of flow solitons that the equation for $\Phi$ can be determined from those of $g$ and $F$ up to a constant -- this constant is significant in a Perelman style treatment of the problem.

Clearly, the fixed points of the flow are described by the equations
\begin{align}
R_{\mu\nu}+ n c F_{\mu\lambda_1\dots \lambda_{n-1}} F_{\nu}{}^{\lambda_1\dots \lambda_{n-1}}-2 a \nabla_\mu\partial_\nu \Phi&=0~,
\label{gradsol1} \\
\nabla^\mu F_{\mu\lambda_1\dots \lambda_{n-1}}+2 a \nabla^\mu \Phi F_{\mu\lambda_1\dots \lambda_{n-1}}&=0~,~~~c\not=0~.
\label{gradsol}
\end{align}
Solutions of these equations are called {\it gradient solitons} with $\Phi$ the gradient scalar  -- in turn the equations are referred to as gradient soliton equations.

On the other hand, the equations that describe the fixed points of the flow up to the natural symmetries of the system, which in this case are diffeomorphisms of $M^D$ generated by a vector field\footnote{ We use $X$ to denote both the vector field and associated 1-form. It should be clear from the context to what $X$ is referred to. Indices are raised and lowered with the metric $g$ as usual.}  $X$ and gauge transformations of $A$ with parameter the $(n-1)$-form  $\Lambda$, are
\begin{align}
R_{\mu\nu}+ n c F_{\mu\lambda_1\dots \lambda_{n-1}} F_{\nu}{}^{\lambda_1\dots \lambda_{n-1}}-2 a \nabla_\mu\partial_\nu \Phi&=\mathcal{L}_X g_{\mu\nu}= \nabla_\mu X_\nu+ \nabla_\nu X_\mu~,
\label{deTruck} \\
2c\Big(\nabla^\mu F_{\mu\lambda_1\dots \lambda_{n-1}}+2 a \nabla^\mu \Phi F_{\mu\lambda_1\dots \lambda_{n-1}}\Big)&=\mathcal{L}_X A_{\lambda_1\dots \lambda_{n-1}}+ d\Lambda'_{\lambda_1\dots \lambda_{n-1}}
\cr
&= X^\mu F_{\mu\lambda_1\dots \lambda_{n-1}}+d\Lambda_{\lambda_1\dots \lambda_{n-1}}~,~~~c\not=0~,
\label{steadysol}
\end{align}
where in the last equality we have used the well-known result that the Lie derivative of the gauge potential $A$ along the vector field $X$ can be rewritten as the inner derivation of the field strength $F$ with respect to $X$ and a gauge transformation of $A$. The solutions to the equations \eqref{deTruck}-(\ref{steadysol}) are referred to as the {\it steady solitons} of the flow (\ref{flow1}).

In the context of 2-dimensional sigma models, where $A$ is identified with the $B$-field, the gradient soliton and steady soliton equations have an interpretation.  The gradient soliton equation is the equation required for the sigma model to be conformally invariant at one loop while the steady soliton equation is the equation required for the sigma model to be scale invariant at one loop. For the more general brane systems that we are considering, there is no such a  worldvolume interpretation. In particular, the flow equations (\ref{flow1}) do not have a renormalisation group flow interpretation for some underlying worldvolume theory. Nevertheless, as we shall demonstrate, the flows exhibit many of the desirable  properties of renormalisation group flow, like monotonicity, using a constrained variation of the brane functional (\ref{dpbrane}), following Perelman's original work.

The steady soliton equations generalise a similar equations that appears in numerical relativity for the Einstein-Maxwell system. In this context the main issue is how to produce equations that are elliptic. The modification of \eqref{deTruck} by the addition of the Lie derivative of a vector goes back to \cite{DeTruck} (a work that was done in the context of Ricci flows) and it is often referred as the DeTruck trick; see \cite{Headrick:2009pv} for an application to numerical relativity. The extension of this method to Einstein-Maxwell system is discussed in \cite{Withers:2014sja, Dias:2015nua, Donos:2015eew}. The steady soliton equations \eqref{deTruck}-\eqref{steadysol} generalise the equations in these papers in a manifestly diffeomorphism invariant way, and we are also able to prove a number of properties, as discussed in this section.

D$p$-brane actions exhibit a non-trivial dilaton field $\Phi$ -- the D$p$-brane solutions exhibit a non-constant dilaton apart from the D3-brane. However, from the perspective of flows, one can consider the flow equation (\ref{flow1}) without the dilaton field, {\it i.e.} set $\Phi=0$ in (\ref{flow1}). Then, one can define
the gradient solitons\footnote{In this definition of the gradient solitons, the emphasis is put on their role as solutions to the field equations of an action, like (\ref{dpbrane}). However, on many occasions in the literature these are defined as a special case of steady solitons upon setting $X=qd\Phi$ in the steady soliton equations. As, we explain in this paragraph, the two definitions can potentially differ. However, it is always possible to make them coincide after a suitable choice of the coupling constants of the action.} as in (\ref{gradsol}) by introducing a dilaton field $\Phi$.  In anticipation of the use of the functional $S$ to investigate the properties of the flow, the couplings of $\Phi$ are fixed as indicated. Then, one can define the steady solitons as in (\ref{steadysol}), with again $\Phi=0$, by requiring that the fixed points of the flow with $\Phi=0$ are satisfied up to the gauge transformations of the system, {\it i.e.} the new steady soliton equation is again given by  (\ref{steadysol}) but with $\Phi=0$. In the case of Ricci solitons as well as that of fundamental string flow solitons, the steady soliton equation implies the gradient soliton equation upon setting $X= q d\Phi$ for some constant $q$. This is not the case for the more general flows (\ref{flow1}), we are considering. This is unless we impose additional restrictions on the constant couplings of $S$ in (\ref{dpbrane}).  In particular, if we set $X= q d\Phi$ and demand that the steady  soliton equations imply the gradient soliton equations, from the Einstein equation, one would deduce that $q=a$. While from the equation for the gauge potential $A$, one would deduce $c=-1/4$. Of course, the $c$ coupling in the action (\ref{dpbrane}) can be adjusted by for example re-scaling the gauge potential $A$.  So it is always possible to produce equations, where the steady soliton equations imply the gradient soliton equations upon requiring that $X$ is given in terms of the gradient of a scalar as $X=q d\Phi$.

Before we proceed further, let us demonstrate that the gradient soliton equation (\ref{gradsol}) for the metric $g$ and gauge potential $A$ imply the field equation for $\Phi$ as derived from (\ref{dpbrane}) up to constant. In sigma models, this is known as the Curci-Paffuti relation \cite{CP}.
 To verify this, the gradient flow equations  for $g$ and $F$ can be rewritten as
\begin{align}
R_{\mu\nu}+ n c F^2_{\mu\nu}-2a \nabla_\mu\nabla_\nu \Phi&=0~,
\cr
\nabla^\mu \big(e^{2a\Phi} F_{\mu\nu_1\dots \nu_{n-1}}\Big)=0~.
\label{gfeqn}
\end{align}
To proceed, we take the divergence  of the first equation to find
\begin{align}
&\nabla^\mu R_{\mu\nu}+ n c \nabla^\mu F_{\mu\lambda_1\dots \lambda_{n-1}} F_{\nu}{}^{\lambda_1\dots \lambda_{n-1}}
\cr
& \qquad\qquad + nc
F_{\mu\lambda_1\dots \lambda_{n-1}} \nabla^\mu F_{\nu}{}^{\lambda_1\dots \lambda_{n-1}}-2a \nabla^2\nabla_\nu\Phi=0~.
\label{dphi}
\end{align}
To evaluate this,   one can establish that
\be
\nabla^2 \nabla_\nu\Phi=\nabla_\nu \nabla^2 \Phi+R_{\nu\lambda} \nabla^\lambda\Phi~,
\ee
after using the usual curvature identity. In addition, the equation for $F$ yields
\be
\nabla^\mu F_{\mu\lambda_1\dots \lambda_{n-1}}=-2a \nabla^\mu\Phi F_{\mu\lambda_1\dots \lambda_{n-1}}~.
\label{divF}
\ee
Moreover, from the Bianchi identity, {\it i.e.} the closure of $F$, one has that
\be
\nabla_\mu F_{\nu\lambda_1\dots \lambda_{n-1}}-\nabla_\nu F_{\mu\lambda_1\dots \lambda_{n-1}}+(n-1) \nabla_{[\lambda_1} F_{|\mu\nu|\lambda_2\dots \lambda_{n-1}]}=0~,
\ee
and so
\be
n F^{\mu\lambda_1\dots \lambda_{n-1}} \nabla_\mu F_{\nu\lambda_1\dots \lambda_{n-1}}=\frac{1}{2} \nabla_\nu F^2~.
\ee
Substituting these into (\ref{dphi}), using the covariant conservation of the Einstein tensor $\nabla^\mu R_{\mu\nu}=1/2 \nabla_\nu R$ and again the Einstein equation in (\ref{gfeqn}), we find that (\ref{dphi}) can be expressed as
\be
\nabla_\nu \Big(R+c F^2-4a \nabla^2 \Phi-4a^2 (\nabla\Phi)^2\Big)=0~.
\label{constdil}
\ee
As we shall demonstrate, the equation in parenthesis is the dilaton field equation as derived by (\ref{dpbrane}) provided that the  constant  $b=4$.  Therefore, the gradient soliton equations imply the field equation for the dilaton up to a constant. It should be noted that this constant is of significance in the theory of sigma models as it is related to the central charge of the theory at the conformal point.

\subsection{Effective brane actions and flows  }\label{preliminaries}

To prove the monotonicity of the flow (\ref{flow1}) and establish a relation between gradient and steady solitons, it is required to construct an effective action for this flow equations.  As an ansatz, we consider the brane action (\ref{dpbrane}) and begin by evaluating the field equations. In particular varying the functional and after a straightforward calculation, we find
\be
\delta S=- \int_{M^D} d^Dx \sqrt{g}\, e^{2a\Phi}\,\Big( (E_{\mu\nu}-\frac{1}{4} P g_{\mu\nu}) \delta g^{\mu\nu} + M_{\lambda_1\dots \lambda_{n-1}} \delta A^{\lambda_1\dots \lambda_{n-1}} + a   P  \delta\Phi\Big)~,
\label{deltaS}
\ee
where
\begin{align}
E_{\mu\nu}&=R_{\mu\nu}+ (b-4) a^2 \partial_\mu\Phi \partial_\nu \Phi-2 a \nabla_\mu\partial_\nu \Phi+ (4-b)\Big( a^2 (\partial \Phi)^2 + \frac{1}{2}  a \nabla^2\Phi \Big) g_{\mu\nu}
\cr
&\quad + n c F_{\mu\lambda_1\dots \lambda_{n=1}} F_\nu{}^{\lambda_1\dots \lambda_{n=1}}~,
\label{eeqn} \\
M_{\lambda_1\dots \lambda_{n-1}}&=-2 ce^{-2a\Phi}\nabla^\mu\big( e^{2a\Phi} F_{\mu\lambda_1\dots \lambda_{n-1}}\big)~,
\label{meqn}
\end{align}
and
\be
P= 2 e^{-a\Phi}\big( -b \nabla^2 +R+ c F^2) e^{a\Phi}=2  \big(-ab \nabla^2 \Phi-b a^2 (\partial\Phi)^2+ R+ c F^2\big)~.
\label{phifeqn}
\ee
It is clear from this that we should set $b=4$ because otherwise  $S$ cannot serve as the effective theory for the flows (\ref{flow1}) as  $\Phi$ does not exhibit the required couplings of a  gradient scalar.  Note that for $b=4$ gives $p=5$ for the couplings as stated in \cite{ksmt}, {\it i.e.} it is the action of a D5-brane background -- perhaps this is not too surprising as D5-branes are dual to the NS5-branes and the latter are a magnetic dual to fundamental strings. Having said that there is no restriction on the degree of the $n$-form $F$ -- one expects for 5-branes $F$ to be a 3-form but this is not the case here.

However, even if one sets $b=4$ in (\ref{dpbrane}), $S$ cannot be used as the effective theory for the flow (\ref{flow1}) because the field equations imply that $P=0$ instead of
$P=\mathrm{const.}$ required by the flow equations in (\ref{constdil}). To accommodate this, $S$ has to undergo a Perelman style modification to be described in the next section below.

Before we proceed with the Perelman style modification,   it should be stressed that for $b\not=4$, {\it i.e.} the value for $b$ that $\Phi$ cannot be the gradient scalar, $S$ can still be used as a starting point to construct an effective theory for a flow. In such a case, $\Phi$ should be thought as another field in the theory instead of the gradient scalar that couples non-linearly. However, this will necessitate a modification of the  flow equations (\ref{flow1}) and a new flow  equation must be added for $\Phi$. This question will be investigated later.

\subsection{Perelman style flows}\label{ssec:three}

Following Perelman \cite{Perelman} we now show that it is possible to modify the brane action (\ref{dpbrane}) such that a suitable volume is preserved along the flow. This is possible due to the universality of the coupling of the $\Phi$ field.
Assuming that $M^D$ is compact without boundary, the brane action can be rewritten as
\be
S=- \int_{M^D}\, d^Dx \,\sqrt{g}\, e^{ a\Phi} \big(R- b \nabla^2+ c F^2\big) e^{ a\Phi}~,
\label{act1}
\ee
  up to a surface term that vanishes upon integration over $M^D$.

We would like to keep the $\Phi$-dependent volume,
\begin{equation}
 \int_{M^D} d^Dx\, \sqrt{g}\, e^{2a\Phi}   \label{vol}
\end{equation}
fixed and equal to one. We comment below on the meaning of this condition in string theory. Following \cite{gpew}, this can be achieved by adding a Lagrange multiplier in the action \eqref{dpbrane}. Extremising this action implies that should impose the condition
\be
C\equiv\int_{M^D} d^Dx \sqrt{g} e^{2a\Phi}-1=0~,
\label{constr}
\ee
and consider the ground state of the operator
\be
(-b \nabla^2+ R+ c F^2) e^{a \Phi}=\lambda\, e^{a \Phi}~,
\label{eing}
\ee
where $\lambda=\lambda(g, F)$ is the eigenvalue of the ground state. This is a Schr\"odinger type of operator and it is known that the ground state is unique and the associated eigenstate is a positive function. This allows to write the eigenstate as the exponential of a smooth function on the manifold as indicated. Note that when \eqref{eing} holds the dilaton field equations valuates to $P=2 \lambda$.

Then, one defines a new  functional
\be
\bar S(g, F)\equiv S(g, F, \Phi(g, F))\vert_{C=0}= -\lambda(g, F)~.
\ee
$\bar S(g, F)$ depends smoothly on $g$ and $F$ as the above equation (\ref{eing}) is elliptic and so the eigenvalues vary smoothly with the parameters $g$ and $F$ of the equation. The functional $\bar S$ is not local, {\it i.e.} it is not written as the integral of a local function of the fields. Instead, it contains some mild non-locality
due to the non-local constraint (\ref{constr}).
Then, observe that for $M^D$ compact without boundary
\begin{align}
\delta \bar S(g, F)&=\delta S(g, F, \Phi)\vert_{C=0, \Phi=\Phi(g, F)}
\cr
&=-\Bigg[\int_{M^D} d^Dx \sqrt{g}\, e^{2a\Phi}\, \Big( E_{\mu\nu} \delta g^{\mu\nu}+ M_{\lambda_1\dots \lambda_{n-1}} \delta A^{\lambda_1\dots \lambda_{n-1}}
\cr
&\qquad\qquad\qquad + P (a \delta\Phi-\frac{1}{4} g_{\mu\nu} \delta g^{\mu\nu})\Big)\Bigg]\Biggl|_{C=0, \Phi=\Phi(g, F)}
\cr
&=-\Bigg[\int_{M^D} d^Dx \sqrt{g}\, e^{2a\Phi}\, \Big( E_{\mu\nu} \delta g^{\mu\nu}+ M_{\lambda_1\dots \lambda_{n-1}} \delta A^{\lambda_1\dots \lambda_{n-1}} \Big)+ \lambda \delta C \Bigg]\Biggl|_{C=0}
\cr
&=-\int_{M^D} d^Dx \sqrt{g}\, e^{2a\Phi}\, \Big( E_{\mu\nu} \delta g^{\mu\nu}+ M_{\lambda_1\dots \lambda_{n-1}} \delta A^{\lambda_1\dots \lambda_{n-1}}\Big)~,
\label{varbarS}
\end{align}
where we have used (\ref{deltaS}), (\ref{eing}) and (\ref{constr}) to establish the first, second and third equalities, respectively.  Furthermore, the field equations $E$ and $M$ are given in (\ref{eeqn}) and (\ref{meqn}), respectively.

It should be emphasised that the whole construction works without any significant change provided that instead of a single $n$-form field strength $c F^2$ in either (\ref{dpbrane}) or (\ref{act1}), one considers instead a term  that contains couplings of several form field strengths of various degrees, {\it i.e.} the term $\sum_n c_n F^2_n$. The computation of the field equations is straightforward and  a Perelman style modification can be still be carried out without a change. Of course, additional flow equations should be added for each additional form gauge potential.

So far, we have not imposed a restriction on the constant $b$. However, in order to explore the properties of the flow equation (\ref{flow1}), we set $b=4$.  In such a case, $\Phi$ becomes the gradient scalar of the flow (\ref{flow1}).  Moreover as we have demonstrated, the gradient flow equations for $g$ and $A$ in (\ref{gradsol}) imply the field equation for $\Phi$ up to a constant.  This is consistent with the eigenvalue equation (\ref{eing}) as the eigenvalue $\lambda$ is related to that constant.  Note though that the constant that appears in the field equation for $\Phi$, as the eigenvalue $\lambda$, depends of the fields $g$ and $F$.

We now comment on the volume \eqref{vol} for the cases that are realised in string theory. As mentioned in the introduction, the $b=4$ case is realised for $p=5$ in $D=10$: the NS5 and D5 branes.
In the NS5 case, $a=-1$ and the dual frame is the standard string frame. The volume in \eqref{vol} is invariant under T-duality transformations. If the spacetime has a U(1) isometry then string theory is invariant under T-duality transformations. In particular, both the sigma model and the low-energy effective action is invariant under them. Using adapted coordinates the spacetime metric may be written in the form,
\begin{equation}
    ds^2 = g_{xx} (dx + A_i dx^i)^2 + \bar{g}_{ij} d x_i d x^j~,
\end{equation}
where $x$ is the isometry direction.
The relevant T-duality rules that needed to verify the invariance of \eqref{vol} are given
\begin{equation}
    g_{xx}'=1/g_{xx}, \quad \bar{g}_{ij}'=\bar{g}_{ij}, \qquad \exp \left(-2 \Phi'\right)=g_{xx} \exp \left(-2 \Phi\right)
\end{equation}
see, for example, \cite{Skenderis:1999bs}, and thus $e^{-2 \Phi} \sqrt{g}$ is invariant. In the case of D5 branes, we have $a=1$, and the volume that we keep fixed is the S-dual of the T-duality volume. Indeed, under S-duality (whose rules are reviewed in \cite{Skenderis:1999bs}), the string frame maps to the dual D1-frame metric, which is relevant frame for D5 branes and the dilaton transforms as $\Phi \to -\Phi$.

\subsection{Monotonicity of the flow}\label{monotone}

Brane flows, as those of (\ref{flow1}), can exhibit a rather involved behaviour. For example, they can be periodic in $t$. For a quantum field theory, such a behaviour of the renormalisation group flow allows for the possibility that starting from a particular energy scale, the theory will not reach either an infrared or an ultraviolet fixed point for all $t$.
One way to rule out such a behaviour is to find a function that either descreases or increases along the flow, {\it i.e.} a monotonic function along the flow.

For the brane flow (\ref{flow1}), we have been investigating, this function is the functional $\bar S$, we have constructed. Indeed set $b=4$ and take the derivative
of $\bar S$ along the brane flow to find that
\begin{align}
\frac{d}{dt} \bar S&=-\int_{M^D} d^Dx\, \sqrt{g}\, e^{2a\Phi}  \big (E_{\mu\nu} \frac{d}{dt} g^{\mu\nu}+ M_{\lambda_1\dots \lambda_{n-1}} \frac{d}{dt}A^{\lambda_1\dots \lambda_{n-1}}\Big)
\cr
&=-\int_{M^D} d^Dx \,\sqrt{g}\, e^{2a\Phi}  \Big (E_{\mu\nu} E^{\mu\nu}+M_{\lambda_1\dots \lambda_{n-1}}M^{\lambda_1\dots \lambda_{n-1}}\Big)\leq 0~.
\end{align}
Therefore $\bar S$ decreases along the flow and so the flow is monotonic.  Moreover, the critical points of the flow are the gradient solitons (\ref{gradsol}).

It should be emphasised though that the monotonicity of the flow does not rule out potential singularities. In geometric theories, these singularities usually manifest themselves as curvature singularities, {\it i.e.} the curvature of the manifold blows up at some finite value of the parameter $t$. Such a behaviour is catastrophic and the flow stops as it cannot continue through the singularity.
In quantum field theory, typically the singularities manifest themselves as strong coupling limits, at either the infrared or ultraviolet. As a result the physics at those points, if it is defined, significantly varies from the physics before the singularity.

\subsection{Gradient solitons from steady solitons}\label{ssec:scaleconf}

Another use of the functional $\bar S$ is to show that all steady solitons (\ref{steadysol}) are in fact gradient solitons (\ref{gradsol}), again for the case $b=4$.
In the context of brane flows, this may not seem significant. However, in the context of sigma models renormalisation group flows, the equations (\ref{steadysol}) describe the conditions for the theory to be scale invariant while the equations that describe the gradient solutions (\ref{gradsol}) are the conditions for the theory to be conformally invariant. A proof that all steady solitons are in fact gradient will mean that all scale invariant sigma models are in fact conformal.

Perhaps, the most appealing part of the proof is its elegance. It is based on the invariance of the functional $\bar S$ under diffeomorphisms of $M^D$ and gauge transformations of the gauge potential $A$.  This is because the brane action $S$ in (\ref{dpbrane}) is diffeomorphic invariant and gauge invariant -- for the latter observe that it depends only on the field strength $F$.  Moreover, all the constraints that are used to construct $\bar S$ from $S$ also are invariant under the same symmetries.

Recall that a gradient solution satisfies $E_{\mu\nu}=M_{\lambda_1\dots \lambda_{n-1}}=0$ (given in \eqref{eeqn}-\eqref{meqn} with $b=4$). Next, suppose that $(g, F)$ is a steady soliton, {\it i.e.} it satisfies the equation
\begin{align}
{\cal E}_{\mu\nu} &\equiv E_{\mu\nu}-{\mathcal L}_X g_{\mu\nu}=E_{\mu\nu}-\nabla_\mu X_\nu-\nabla_\nu X_\mu=0~,
\cr
{\cal M}_{\lambda_1\dots \lambda_{n-1}}&\equiv M_{\lambda_1\dots \lambda_{n-1}}-{\mathcal L}_X A_{\lambda_1\dots \lambda_{n-1}}- (d\Lambda')_{\lambda_1\dots\lambda_{n-1}}
\cr
&=M_{\lambda_1\dots \lambda_{n-1}}-X^\mu F_{\mu\lambda_1\dots \lambda_{n-1}}- (d\Lambda)_{\lambda_1\dots \lambda_{n-1}}=0~,
\end{align}
for some vector field $X$ and a gauge transformation with parameter $\Lambda$.  As $\bar S$ is invariant under diffeomorphisms of $M^D$ and the gauge transformations of $A$,  this implies that $\delta \bar S=0$ provided that the metric $g$ and $F$ transform as
\be
\delta g_{\mu\nu}=\nabla_\mu X_\nu+\nabla_\nu X_\mu~,~~~\delta A_{\lambda_1\dots \lambda_{n-1}}= X^\mu F_{\mu\lambda_1\dots \lambda_{n-1}}+ (d\Lambda)_{\lambda_1\dots \lambda_{n-1}}\, ,
\ee
with $g$ and $F$ arbitrary, {\it i.e.} under these variations,
\begin{equation} \label{gauge_inv}
    0=\delta \bar S= \int_{M^D} d^Dx\, \sqrt{g}\, e^{2a\Phi}  \big (E_{\mu\nu} \delta g^{\mu\nu}+ M_{\lambda_1\dots \lambda_{n-1}} \delta A^{\lambda_1\dots \lambda_{n-1}}\Big)
\end{equation}
Then,
\begin{align}
\frac{d}{dt} \bar S&=-\int_{M^D} d^Dx\, \sqrt{g}\, e^{2a\Phi}  \big (E_{\mu\nu} \frac{d}{dt} g^{\mu\nu}+ M_{\lambda_1\dots \lambda_{n-1}} \frac{d}{dt}A^{\lambda_1\dots \lambda_{n-1}}\Big)
\nonumber \\
=&-\int_{M^D} d^nx\, \sqrt{g}\, e^{2a\Phi} \Big( E^{\mu\nu} {\cal E}_{\mu\nu}+{M}^{\lambda_1\dots \lambda_{n-1}} {\cal M}_{\lambda_1\dots \lambda_{n-1}} \Big)\, ,
\end{align}
and for a steady soliton, ${\cal E}={\cal M}=0$,
\begin{align}
0=&\int_{M^D} d^nx\, \sqrt{g}\, e^{2a\Phi} \Big( E^{\mu\nu} {\cal E}_{\mu\nu} + {M}^{\lambda_1\dots \lambda_{n-1}} {\cal M}_{\lambda_1\dots \lambda_{n-1}}\Big)
\cr
=&\int_{M^D} d^nx\, \sqrt{g}\, e^{2a\Phi} \Big({E}_{\mu\nu} E^{\mu\nu}+{M}_{\lambda_1\dots \lambda_{n-1}} {M}^{\lambda_1\dots \lambda_{n-1}}\Big)~,
\end{align}
where
 the cross term vanishes because of \eqref{gauge_inv}. Thus, $E=M=0$ and all steady solitons are gradient solitons. For the proof, it is sufficient to assume that $M^D$ is compact without boundary  and the fields are smooth.

This theorem has some consequences. Let us choose the couplings in the action (\ref{dpbrane}) such that  steady soliton equations (\ref{steadysol}) (with $\Phi=0$), {\it i.e.},
\begin{align}
R_{\mu\nu}-\frac {n}{4}  F_{\mu\lambda_1\dots \lambda_{n-1}} F_{\nu}{}^{\lambda_1\dots \lambda_{n-1}}&= \nabla_\mu X_\nu+ \nabla_\nu X_\mu~,
\cr
-\frac{1}{2}\nabla^\mu F_{\mu\lambda_1\dots \lambda_{n-1}}
&= X^\mu F_{\mu\lambda_1\dots \lambda_{n-1}}+d\Lambda_{\lambda_1\dots\lambda_{n-1}}~,~~~
\label{steadysol2}
\end{align}
reduce to gradient soliton equations (\ref{gradsol})  upon setting  $X=a d\Phi$. We have seen that in such a case  $c=-1/4$. Suppose that we have a solution to the steady soliton equation above.  The theorem we have demonstrated implies that it must also be a solution of the gradient flow equations (\ref{gradsol}). This implies that there are two possibilities. One is that $X$ is the gradient of a scalar, $X=a d\Phi$. Alternatively, there must exist a Killing vector field $Y$ with
\be
Y=X-a d\Phi~,
\ee
such that $\iota_Y F=-d\Lambda$, which in turn implies that  $F$ is also  invariant, $\mathcal{L}_Y F=0$.  Therefore, steady  solutions  either admit a vector field $X$ that it is the gradient of a scalar or they admit  the action of a vector field $Y$ that leaves invariant both the metric $g$ and form field strength $F$.

Moreover, $Y$ leaves the gradient scalar $\Phi$ also invariant.  To see this, recall that $P$ in (\ref{phifeqn}) is constant.  Taking the Lie derivative of $P$ and using that $Y$ is Killing and leaves $F$ invariant, we find that
\be
 \nabla^2 \mathcal{L}_Y\Phi+2a  g^{\mu\nu}\partial_\mu\Phi \partial_\nu\mathcal{L}_Y\Phi =0~.
\ee
As $M^D$ is compact without boundary upon applying the maximum principle, one concludes that $\mathcal{L}_Y\Phi$ is constant. However, as $\mathcal{L}_Y\Phi=Y^\mu\partial_\mu\Phi$, an integration over $M^D$ implies that the constant must vanish and so $\mathcal{L}_Y\Phi=0$.

\subsection{Gradient solitons}

A special property of the gradient equations (\ref{gradsol}) (when $D=10, n=3$) is that they are the field equations of the bosonic sector of II supergravities (in the dual frame \footnote{The terminology dual frame has the following meaning. Each $p$-brane couples electrically to a $(p+1)$ potential and the corresponding (Hodge) dual field strength is an $(8-p)$ form (in ten dimensions). In the dual frame this field strength and the graviton couple of the dilaton the same way, {\it i.e.} the effective action takes the form \eqref{dpbrane}. For example, for the cases relevant to us: the dual frame for NS5 branes is the standard string frame, and the dual frame for D5 branes is the D1-frame.
The dual frame was originally introduced in \cite{Duff:1990hb, Duff:1991pe,Duff:1994fg}, and solutions and their properties in this frame were discussed \cite{Boonstra:1998mp, ksmt}.}). This then implies that the 5-brane solutions are examples of gradient solitons.

The 5-brane solutions in the dual frame are given by \cite{Duff:1990wv, Callan:1991dj, Horowitz:1991cd, Duff:1994an}:
\begin{align} \label{5brane}
    ds^2 &= -dt^2 + dx_1^2 + \cdots +dx_5^2 + H  ds^2(\mathbb{R}^{4}) \\
    F_{(3)} &= \star d H \label{3-form} \\
    \exp \Phi &= H^{\pm 1/2} \label{dilaton}
\end{align}
where $ds^2(\mathbb{R}^{4})$ is the standard Euclidean metric on $\mathbb{R}^4$ and $H$ is a harmonic function on $\mathbb{R}^4$,
\begin{equation}
    \nabla^2 H =0\, ,
\end{equation}
where $\nabla^2$ is the Laplacian on $\mathbb{R}^4$,
and $\star$ is the Hodge dual in the transverse $\mathbb{R}^4$. The solution with the plus sign in \eqref{dilaton} is the NS5 brane solution and the one with the minus sign the $D5$ brane solution. The NS5 brane solution  is a solution of both IIA and IIB supergravity.
The fact that there are two solutions differing in the sign of $\Phi$ is due to the S-duality of IIB supergravity.
The NS5 brane is a solution in the string frame and the corresponding gradient equations are given by (\ref{gradsol}) with $D=10, n=3, a=-1$. The 3-form in \eqref{3-form} is the field strength of the Kalb-Ramond NS-NS antisymmetric tensor.
The $D5$ brane is a solution in the $D1$-brane frame, and the corresponding gradient equations are given by (\ref{gradsol}) with $D=10, n=3, a=1$. The 3-form in \eqref{3-form} is the field strength of the R-R antisymmetric tensor.

In the physical context, the harmonic function is usually taken to be equal to
\begin{equation} \label{H1}
    H = 1 + \frac{Q}{r^2}\,,
\end{equation}
where $r$ is the polar radial coordinate on $\mathbb{R}^4$, the constant in $H$ was set equal to 1 so that the solution is asymptotically flat and $Q$ is a constant related to the number of 5-branes. Other solutions include multi-center solutions,
\begin{equation} \label{multi}
    H = 1 + \sum_i \frac{Q_i}{|\vec{x}-\vec{x}_i|^2}
\end{equation}
where $\vec{x} \in \mathbb{R}^4$ and $\vec{x}_i$ are the positions of the 5-branes, and solutions that involve more general harmonic functions,
\begin{equation} \label{H2}
    H = \sum_{r,k} \left( l_{k I} r^{k} + \frac{h_{k I}}{r^{k +2}}\right) Y_k^I
\end{equation}
where $Y_k^I$ are spherical harmonics on $S^3$ and $l_{k I}, h_{k I}$ are constants. These solutions are extermal and supersymmetric. A closely related solution may be obtained by reversing the sign in $F_{(3)}$:  these solutions represent the anti-branes corresponding to the original branes.

A distinguished special case is to consider the near-horizon $r \to 0$ limit of the standard 5-branes based on \eqref{H1}. In this limit, the solution simplifies to \cite{Gibbons:1993sv, Boonstra:1998mp}
\begin{align} \label{5brane_NH}
    ds^2 &= -dt^2 + dx_1^2 + \cdots +dx_5^2 + d y^2  + Q d \Omega_3^2 \\
    F_{(3)} &=  2 Q vol (S^3) \label{3-form} \\
    \Phi &= \mp y  \label{dilaton}
\end{align}
where $y = \log r$,  $d \Omega^2_3$ is the standard metric on the unit 3-sphere and $vol (S^3)$ is the volume form on $S^3$. The metric is (locally) 7-dimensional Minkowski spacetime times a 3-sphere with a linear dilaton. The linear dilaton may be invariantly characherised as the  Killing potential for the Killing vector associated with translational invariance in the $y$ direction.

In our context, we are interested in Euclidean solutions and on a compact manifold. Since the above solutions are static, they can be Wick rotated to Euclidean signature by setting $t=-i \tau$. If the transverse geometry is compact, call it ${\cal M}^4$ (for example, a torus), then the harmonic function $H$ must be equal to a constant, which then implies that the metric is (locally) $\mathbb{R}^{6} \times {\cal M}^4$, the 3-form vanishes and dilaton is a constant. Note also that that for the 5-branes solutions, $\sqrt{g} e^{2 a \Phi} = H$, and the volume \eqref{vol}
is infinite if the spacetime is non-compact.
However, one may wish to relax the condition of the compactness and replace them by suitable fall off conditions and require completeness. The metric in \eqref{5brane} with $H$ in \eqref{H1} is indeed complete: as $r \to 0$ the geometry becomes that of \eqref{5brane_NH}, which is indeed a smooth geometry. 
We note however, that as the dilaton grows the system becomes strongly coupling regime, see
\cite{Itzhaki:1998dd, Aharony:1998ub} for relevant discussions.

It would be interesting to investigate the entire brane flow and see under which conditions the above gradient solutions would appear as actually fixed points. Prior experience with geometric flows, however, suggest that typically the flow will hit a singularity before it arrives at the nice fixed point.

\section{General brane flows}

We have demonstrated that the use of the action (\ref{dpbrane}) to investigate the properties of the flow (\ref{flow1}) with $\Phi$ identified as a gradient scalar requires that the coupling $b=4$. Such a coupling is associated with the D5-branes. If one wishes to generalise the construction to $b\not=4$, $\Phi$ cannot be thought of as a gradient scalar but instead as another field like $g$ and $F$, which couples non-linearly to other fields. Therefore, further progress will depend on the introduction of additional scalar fields in the theory  that can be used to construct a gradient scalar for  the flow. The aim is to maintain the role of an action to demonstrate the properties of the flow, like for example monotonicity.

This can be  achieved with  two adjustments. One is to generalise the action (\ref{dpbrane}) and the other is to modify the flow equations (\ref{flow1}).
A generalisation of  the action (\ref{dpbrane}) can be achieved by introducing new fields  in such a way that the field equations yield those of a gradient brane soliton after possibly a field redefinition. From this, one can proceed to construct brane flow equations. In this construction, the gradient scalar and flow equations are  identified posteriori

An alternative way is to postulate a set of brane flow equations with a given gradient scalar. Then, one can proceed to construct a brane-like action such that the field equations yield the equations of gradient brane solitons of the flow. In this construction, the flows and the gradient scalar are constructed {\it a priori}. Below, we shall pursue both ways of constructing flows and their corresponding actions.

\subsection{Dilaton-axion brane flow}\label{ssec:axion}

Let us begin with the first method that requires a modification of the action (\ref{dpbrane}). A minimal modification that one can make  is to add another scalar field $\Psi'$ as follows:
\begin{align}
S&=- \int_{M^D} d^Dx \,\sqrt{g}\, e^{2a\Phi'} e^{2\alpha\Psi'} \big(R+a^2 b (\partial\Phi')^2+\beta \alpha^2 (\partial \Psi')^2+ c F^2\big)~,
\label{act2}
\end{align}
where $\alpha, \beta$ are new coupling constants and the rest of the fields are as in (\ref{dpbrane}). Neither $\Phi'$ nor $\Psi'$ can be identified with a gradient scalar.  However, let us consider the field redefinition
\be
\Phi=a\,\Phi'+\alpha\, \Psi'~,~~\Psi=\alpha\, \Phi'-a\, \Psi'~.
\ee
 The action (\ref{act2})  in terms of the fields $\Phi$ and $\Psi$ can be re-written as
\begin{align}
S=&- \int_{M^D} d^Dx \,\sqrt{g}\, e^{2\Phi} \Big( R+\frac{1}{(a^2+\alpha^2)^2}\big(a^2 b (a \partial\Phi+\alpha \partial \Psi)^2+ \beta \alpha^2
(\alpha\partial\Phi-a \partial \Psi)^2\big) +c F^2\Big)~.
\end{align}
Aiming for $\Phi$ to be identified as a gradient field, the mixed derivative term between $\Phi$ and $\Psi$ should vanish.  This yields the condition
\be
a^2 b=\beta \alpha^2~.
\label{cond1}
\ee
Using  this, the action can be expressed as
\begin{align}
S=&- \int_{M^D} d^Dx \sqrt{g}\, e^{2\Phi}\, \Big( R+\frac{2 a^2 b}{a^2+\alpha^2} \Big(( \partial\Phi)^2+
( \partial \Psi)^2\Big) +c F^2\Big)~.
\label{act2a}
\end{align}
Therefore,   $\Phi$ can be identified with the dilaton and $\Psi$ with an axion field that has 1-form field strength  $F^{(1)}=d\Psi$.
It is clear that the coupling of $\Psi$ is as that of a gauge field with a 1-form field strength, {\it i.e.} it is as that of the $n$-form field strength $F$.
Because of this, we can repeat the argument used for the action (\ref{dpbrane}) to identify the gradient scalar but now for the dilaton $\Phi$. This yields the condition
\be
a^2 b= 4 (a^2+\alpha^2)~.
\label{cond2}
\ee
for the coupling of $\Phi$.

Therefore, there are two conditions on the couplings given by (\ref{cond1}) and (\ref{cond2}).  These can be solved for $\alpha$ and $\beta$, provided that $b\not=4$,  to find
\be \label{alphabeta}
\alpha^2=\frac{1}{4} a^2 (b-4)~,~~~\beta=\frac{4b}{b-4}~.
\ee
 Thus,  the restriction on the original couplings $a$ and $b$ of the theory becomes $b>4$ for the coupling $\alpha$, and so the action, to be a real.

The brane flow associated with the new action  (\ref{act2a}) is
\begin{align}
\frac{d}{dt} g_{\mu\nu}&=-\big(R_{\mu\nu}+ 4 \partial_\mu\Psi \partial_\nu \Psi+ n c F_{\mu\lambda_1\dots \lambda_{n-1}} F_{\nu}{}^{\lambda_1\dots \lambda_{n-1}}-2  \nabla_\mu\partial_\nu \Phi\big)~,
\cr
\frac{d}{dt} A_{\lambda_1\dots \lambda_{n-1}}&= -2c \Big(\nabla^\mu F_{\mu\lambda_1\dots \lambda_{n-1}}+2  \nabla^\mu \Phi F_{\mu\lambda_1\dots \lambda_{n-1}}\Big)~,
\cr
\frac{d}{dt} \Psi&=- 16 \Big(\nabla^2\Psi+2  \nabla^\mu \Phi \nabla_\mu \Psi\Big)~,
\label{flow2}
\end{align}
where we have introduce a flow equation for the axion $\Psi$ and $\Phi$ is the gradient scalar as expected.

Observe that setting $\Psi=0$ is a consistent solution. The remaining equations could describe a D$p$-brane flow with $b>4$. However, this consistent truncation only describes D5-branes. This follows from the coefficient of $\Phi$ in $e^{2 \Phi}$. Comparing the corresponding coupling in  D$p$-brane actions in \cite{Boonstra:1998mp, ksmt} yields
\be
2 = \frac{2 (p-3)}{7-p} \quad \Rightarrow \quad p=5.
\ee

One can proceed from this to perform a Perelman style modification for the action (\ref{act2a}) and show that the flow (\ref{flow2}) is monotonic. It can also be shown that all steady brane flow solitons of (\ref{flow2}) are gradient brane flow solitons. One of the consequences of the latter is, under the same conditions as those described in section \ref{ssec:scaleconf},  the existence of a Killing vector field $Y$ such that  $Y=X-d\Phi$, where $X$ is the steady soliton vector field. One can also show that $Y$  also leaves  the axion field strength $F^{(1)}$, the $n$-form field strength $F$ and $\Phi$ invariant.

\subsection {An alternative method}

Suppose we have a theory with fields a metric $g$ and a $(n-1)$-form gauge potential $A$.  We wish to construct a flow that generalises that of (\ref{flow1}) for couplings $b\not=4$ for some pre-determined gradient flow scalar $\Phi$.
An alternative approach, to that explained in the previous section,  to generalise the brane flows to $b\not=4$ is to insist that $\Phi$ is a gradient scalar. This is equivalent to stating that the flow for the metric $g$ and $(n-1)$-form gauge potential $A$ is given by

\begin{align}
\frac{d}{dt} g_{\mu\nu}&=- (E_{\mu\nu}+p \nabla_\mu\nabla_\nu \Phi)~,
\cr
\frac{d}{dt} A_{\lambda_1\dots \lambda_{n-1}}&=  \Big(M_{\lambda_1\dots\lambda_{n-1}})+ q \nabla^\mu \Phi F_{\mu\lambda_1\dots \lambda_{n-1}}\Big)
\label{flow3}
\end{align}
where $p,q$ are constants, and $E$ and $M$ is an Einstein equation and the field equation for the gauge potential $A$, respectively, to be determined later.

As we wish to be able to perform a Perelman style modification to the action, the minimal requirement is that $e^{\alpha\Phi}$ couples linearly in the action. Introducing another field $\Psi$, which can be thought of as the ``dilaton'',  an ansatz for an action that generalises (\ref{dpbrane}) is
\begin{align}
S&=- \int_{M^D} d^Dx \sqrt{g} e^{2a\Psi} e^{2\alpha\Phi} \big(R+a^2 b (\partial\Psi)^2+\beta \alpha^2 (\partial \Phi)^2+4a \alpha\gamma \partial\Psi\cdot \partial\Phi+ c F^2\big)
\cr
&=- \int_{M^D} d^Dx \sqrt{g} \Bigg( e^{ 2a\Psi} e^{2\alpha\Phi} \big(R +c F^2\big)+     b  e^{2\alpha\Phi} (\partial e^{a\Psi})^2+ \beta e^{2a \Psi} (\partial e^{\alpha\Phi})^2
\cr
&\qquad\qquad + \gamma \partial e^{2a\Psi}\cdot \partial e^{2\alpha\Phi} \Bigg )
\cr
&==- \int_{M^D} d^Dx \sqrt{g} e^{\alpha\Phi}\Bigg( e^{ 2a\Psi}  \big(R +c F^2\big)+     b  (\partial e^{a\Psi})^2- \beta\nabla^\mu (e^{2a \Psi} \partial_\mu)
\cr
&\qquad\qquad - \gamma \nabla^2 e^{2a\Psi} \Bigg )e^{\alpha\Phi}~.
\label{act3}
\end{align}

After a re-arrangement of terms, the field equation for the gradient scalar $\Phi$, or equivalently for $e^{\alpha\Phi}$, is
\begin{align}
&\Bigg(- \beta\nabla^\mu (e^{2a \Psi} \partial_\mu)+  e^{ 2a\Psi}  \big(R +c F^2\big)+     b  (\partial e^{a\Psi})^2
 - \gamma \nabla^2 e^{2a\Psi} \Bigg )e^{\alpha\Phi}=0~.
\label{gscalar}
\end{align}
In anticipation to the Perelman modification, observe that it is of Strum-Liouville type of equation on $e^{\alpha\Phi}$. It is apparently linear and as $e^{2a\Psi}>0$, it is elliptic. Moreover for   $\beta>0$, it is  in the standard form. It turns out that such a differential equation on compact manifolds has a unique ground state with eigenfunction a strictly positive function on the manifold.

There are additional restrictions on the action for $\Phi$ to be a gradient scalar.  For example, in the Einstein equation there must not be terms like $\partial_\mu \Phi \partial_\nu\Phi$ and $\partial_\mu\Psi \partial_\nu\Phi$. Indeed, varying the action with the metric and upon using (\ref{gscalar}), we find that
\be
E_{\mu\nu}= R_{\mu\nu}+nc F_{\mu\lambda_1\dots \lambda_{n-1}} F_\nu{}^{\lambda_1\dots\lambda_{n-1}}+ (b-4) a^2 \partial_\mu\Psi\partial_\nu\Psi- 2 \nabla_\mu\nabla_\nu( a\Psi+\alpha \Phi )
\ee
provided that
\be
\beta=4~,~~~\gamma=2~.
\label{constv}
\ee
Furthermore, the field equation for $A$ is
\be
 \nabla^\mu F_{\mu\lambda_1\dots \lambda_{n-1}} + \nabla^\mu (a \Psi+\alpha \Phi) F_{\mu\lambda_1\dots \lambda_{n-1}}=0~.
 \ee
 Again, it is linear  in $\partial \Phi$, which is required for $\Phi$ to be a gradient scalar, without any additional restrictions on the couplings.

Next, let us turn to the field equation for the dilaton $\Psi$. Upon using the equation for the gradient scalar (\ref{gscalar}), the field equation for $\Psi$ becomes
\be
P=a (4-b)\big( \nabla^2\Psi+2 a(\partial\Psi)^2\big)+ 6 a \alpha \partial^\mu\Psi \partial_\mu\Phi~.
\ee
Again this is of the form required for $\Phi$ to be a gradient scalar, i.e. it is linear in $\partial \Phi$, without any additional restriction on the couplings.

It is clear that up to a redefinition of the fields and appropriate identification of the coupling constants, these field equations are the same as those of the
action (\ref{act2a}). Indeed setting,
\be
\tilde \Phi=a \Psi+\alpha \Phi~,~~~\tilde \Psi=\Psi~,
\label{frdef}
\ee
the all field equations above assume the form of those of (\ref{act2a}).  The same applies for the action (\ref{act3}) and (\ref{act2a}). After re-defining the fields as (\ref{frdef}),  the action  (\ref{act3}) takes the form  of (\ref{act3}) provided that the coupling constants of the latter take the values (\ref{constv}). This demonstrates that the two approaches taken to generalise the brane flows for $b\not=4$ lead to the same result.

One can proceed to prove the monotonicity of the flow and demonstrate that all steady solitons are in fact gradient solitons.  This can either be done using the action of the previous section  (\ref{act2a}) or pursue these questions using the action (\ref{act3}). Both lead to the same result.  However, in the latter case
to perform a Perelman style of modification, one has to impose the constraint
\be
C=\int_M\, d^Dx\, \sqrt{g}\, e^{2a\Psi+2\alpha \Phi}-1=0~,
\ee
instead of (\ref{constr}).

\section{M-flows}

So far our analysis has been based on actions without  Chern-Simons style of terms. However, if one considers actions like that of 11-dimensional supergravity to perform a similar analysis, a Chern-Simons term has to be added. Apart from that another question can be raised regarding the role of a gradient scalar $\Phi$. Famously, the fields of 11-dimensional supergravity do not contain a scalar field. Nevertheless, in the investigation of flows, even for theories that do not contain a scalar field,  a scalar field $\Phi$ can be introduced. This is because steady solitons associated to a vector field $X$, and so gradient solitons, have a dual interpretation. One is as solutions of equations like (\ref{steadysol}) and the other is as solutions of the flow equations, like (\ref{flow1}) with $\Phi=0$, which are now generated by the flow of the vector field $X$.  This dual interpretation of the flow solitons has been emphasised in {\it e.g.} \cite{BCDK} and applied in the context of sigma model renormalisation group flows in \cite{gp3}. So even though, the flow equations may not contain a vector field $X$, and perhaps a gradient scalar $\Phi$, nevertheless they can be introduced in the context. Thus if one wishes to examine the properties of the flows, a gradient scalar can be introduced even though it does not exist in the theory that the flow is modelled on.

\subsection{Action for M-flows}

Motivated by the introduction above, let us introduce a gradient scalar $\Phi$ from the beginning and continue to add a Chern-Simons style of term loosely based on the properties of the bosonic part of the action of 11-dimensional supergravity. Suppose again that the fields of the theory is a metric $g$ and a $n$-form field strength, $F=dA$. An action with the required properties is
\be
S_T=S+S_{\mathrm{CS}}~,~~~ S_{\mathrm{CS}}=e\, \int_M\, d^Dx\,  A\wedge^k F~,
\label{maction}
\ee
where $S$ is the action given in (\ref{act1}),  $e$ is another coupling constant and $\wedge^k$ denotes the wedge product of $k$ number of $n$-form field strengths $F$.
Clearly $F$, $k>1$, has to be an even-form, otherwise the $S_{\mathrm{CS}}=0$,  and $(k+1)n-1=D$. The action (\ref{maction}) can be easily extended to include more involved Chern-Simons style of terms provided that $F$ in (\ref{act1}) is a multi-form. Most of our considerations below generalize to this case as well.  However to keep our presentation concise,   we shall only investigate in detail the case that $F$ is a $n$-form.

For the action $S_{\mathrm{CS}}$ to be globally defined on $M$, the form $A\wedge^kF$ has to be globally defined. This is particularly essential for proving the relation between steady and gradient solitons below. This is a stronger restriction imposed on Chern-Simons type terms than the usual condition for $A\wedge^kF$ to represent a class in $H^D(M, \bZ)$ -- the latter arises from the weaker requirement that $\exp i S_{\mathrm{CS}}$ must be well-defined instead of $S_{\mathrm{CS}}$.  In appendix A, we give some more  details to justify this restriction.

As $S_{\mathrm{CS}}$ does not depend on either the metric $g$ or the scalar $\Phi$, the field equations for $g$ and $\Phi$ are already been given in (\ref{eeqn})
and (\ref{phifeqn}), respectively.  The only equation that gets affected by the additional term is the field equation for the $(n-1)$-form  gauge potential $A$ that now reads
\be
M_{\lambda_1\dots \lambda_{n-1}}=-2 c\,e^{-2a\Phi}\nabla^\mu\big( e^{2a\Phi} F_{\mu\lambda_1\dots \lambda_{n-1}}\big)+e\, (k+1)\,e^{-2a\Phi}\, {}^*(\wedge^k F)_{\lambda_1\dots \lambda_{n-1}}~,
\label{meqn2}
\ee
where the star operation denotes Hodge duality.

The variation of the action $S_T$ reads as
\be
\delta S_T=- \int_{M^D} d^Dx \sqrt{g}\, e^{2a\Phi}\,\Big( (E_{\mu\nu}-\frac{1}{4} P g_{\mu\nu}) \delta g^{\mu\nu} + M_{\lambda_1\dots \lambda_{n-1}} \delta A^{\lambda_1\dots \lambda_{n-1}} + a   P  \delta\Phi\Big)~,
\label{deltaS2}
\ee
where $E$ and $P$ are given in (\ref{eeqn}) and (\ref{phifeqn}), respectively,  while $M$ is given in (\ref{meqn2}).

The flow equations that we shall investigate in relation to the action above are
\begin{align}
\frac{d}{dt} g_{\mu\nu}&=-\big(R_{\mu\nu}+ n c F_{\mu\lambda_1\dots \lambda_{n-1}} F_{\nu}{}^{\lambda_1\dots \lambda_{n-1}}-2 a \nabla_\mu\partial_\nu \Phi\big)
\cr
\frac{d}{dt} A_{\lambda_1\dots \lambda_{n-1}}&= -2c \Big(\nabla^\mu F_{\mu\lambda_1\dots \lambda_{n-1}}+2 a \nabla^\mu \Phi F_{\mu\lambda_1\dots \lambda_{n-1}}\Big)+e\,(k+1)\,e^{-2a\Phi}\, {}^*(\wedge^k F)_{\lambda_1\dots \lambda_{n-1}}~.
\label{flow3}
\end{align}
Notice the modification of the flow equation for $A$ due Chern-Simons style of term.
It is clear that $\Phi$ is a gradient scalar. Therefore to match the flow above to the Einstein equation (\ref{eeqn}), we again require that $b=4$.

As before, we have not introduced a flow for the gradient scalar $\Phi$.  This is because the field equation for $\Phi$ is implied from those of $g$ and $A$. The calculation for this is similar to that presented in section \ref{ssec:one}. The only difference is that upon evaluating $\nabla^\mu E_{\mu\nu}$, we have to consider the contribution made by the Chern-Simons term. This only enters in the calculation of the term
\be
n\, c\,\nabla^\mu F_{\mu\lambda_1\dots \lambda_{n-1}} F_\nu{}^{\lambda_1\dots \lambda_{n-1}}~.
\ee
Using the field equation for $F$ in (\ref{meqn2}), the contribution that contains $\partial\Phi$ is as in (\ref{divF}) while the Chern-Simons term gives
\be
e\,\frac{n}{2} \,e^{-2a\Phi}\, {}^* (\wedge^k F)_{\lambda_1\dots \lambda_{n-1}}  F_\nu{}^{\lambda_1\dots \lambda_{n-1}}=0~.
\ee
As indicated this expression vanishes because  its Hodge dual is proportional to $(k+1) \iota_X F\wedge^k F=\iota_X (\wedge^{k+1} F)$ and $\wedge^{k+1} F=0$ as it is a degree $D+1$-form on a $D$-dimensional manifold $M^D$, where $\iota_X$ is the inner derivation along any vector field $X$. The rest of the computation remains unchanged and yields that the field equation for $\Phi$ is satisfied up to a constant.

\subsection{Monotonicity of the M-flow}

The proof of the monotonicity of the flow requires to find a functional $\bar S_T$ that is monotonic along the flow (\ref{flow3}).  Let us set the coupling $b=4$ -- we shall comment for the cases that $b\not=4$ later. One way to construct such a functional  is to perform a Perelman style of modification to the action $S_T$.  This involves to imposing the condition
\be
C\equiv\int_{M^D} d^Dx\, \sqrt{g}\, e^{2a\Phi}-1=0~,
\label{constr2}
\ee
and to considering the ground state of the eigenvalue problem
\be
(-b \nabla^2+ R+ c F^2)\, e^{a \Phi}=\lambda\, e^{a \Phi}~,
\label{eing2}
\ee
where $\lambda=\lambda(g, F)$ is the eigenvalue of the ground state. These equations are the same as those in (\ref{constr2}) and (\ref{eing2}) and so have the same properties, {\it e.g.} the ground eigenstate is positive and it can be written as the exponential of a function as indicated.  If $\Phi=\Phi(g, F)$ is the ground eigenstate of the eigenvalue problem (\ref{eing2}), then $\bar S_T$ is defined
as
\be
\bar S_T(g, A)= S_T(g, A, \Phi)\vert_{\Phi=\Phi(g, F),\, C=0}=-\lambda(g, F)+S_{\mathrm{CS}}(A)~.
\ee
Note that $\bar S_T$ is not given in terms of the eigenvalue $\lambda$ of the ground state, as in section \ref{ssec:three}, but now it is corrected by the Chern-Simons type term in the action. Because of the presence of $S_{\mathrm{CS}}(A)$, one may raise the question whether $\bar S_T$ is well-defined. We shall deal with this question below in section \ref{ssec:mgs} below. So, let us assume that it is and proceed with the proof of the monotonicity of the flow.

To prove the monotonicity of the flow, a calculation similar to that presented in (\ref{varbarS}) gives that
\be
\delta \bar S_T=-\int_{M^D} d^Dx \sqrt{g}\, e^{2a\Phi}\, \Big( E_{\mu\nu} \delta g^{\mu\nu}+ M_{\lambda_1\dots \lambda_{n-1}} \delta A^{\lambda_1\dots \lambda_{n-1}}\Big)~,
\ee
where $E$ is given in (\ref{eeqn}) with $b=4$ and $M$ is given in (\ref{meqn2}).  Thus taking the derivative of $\bar S_T$ along the flow gives
\begin{align}
\frac{d}{dt} \bar S_T&=-\int_{M^D} d^Dx \sqrt{g}\, e^{2a\Phi}\, \Big( E_{\mu\nu} \frac{d}{dt} g^{\mu\nu}+ M_{\lambda_1\dots \lambda_{n-1}} \frac{d}{dt} A^{\lambda_1\dots \lambda_{n-1}}\Big)
\cr
&=-\int_{M^D} d^Dx \sqrt{g}\, e^{2a\Phi}\, \Big( E_{\mu\nu} E^{\mu\nu}+ M_{\lambda_1\dots \lambda_{n-1}} M^{\lambda_1\dots \lambda_{n-1}}\Big)\leq 0~.
\end{align}
Thus $\bar S_T$ decreases along the flow and so the M-flow (\ref{flow3}) is monotonic.

\subsection{Gradient solitons  from steady solitons for M-flows}\label{ssec:mgs}

The gradient solitons are the fixed points of the M-flow and satisfy the equations
\begin{align}
R_{\mu\nu}+ n c F_{\mu\lambda_1\dots \lambda_{n-1}} F_{\nu}{}^{\lambda_1\dots \lambda_{n-1}}-2 a \nabla_\mu\partial_\nu \Phi&=0~,
\cr
 2c \Big(\nabla^\mu F_{\mu\lambda_1\dots \lambda_{n-1}}+2 a \nabla^\mu \Phi F_{\mu\lambda_1\dots \lambda_{n-1}}\Big)-e\, (k+1) \,e^{-2a\Phi}\, {}^*(\wedge^k F)_{\lambda_1\dots \lambda_{n-1}}&=0~.
\label{grad3}
\end{align}
While the steady solitons satisfy the equation,
\begin{align}
R_{\mu\nu}+ n c F_{\mu\lambda_1\dots \lambda_{n-1}} F_{\nu}{}^{\lambda_1\dots \lambda_{n-1}}-2 a \nabla_\mu\partial_\nu \Phi&=\nabla_\mu X_\mu+\nabla_\nu X_\mu~,
\cr
 2c \Big(\nabla^\mu F_{\mu\lambda_1\dots \lambda_{n-1}}+2 a \nabla^\mu \Phi F_{\mu\lambda_1\dots \lambda_{n-1}}\Big)-e\, (k+1) \,e^{-2a\Phi}\, {}^*(\wedge^k F)_{\lambda_1\dots \lambda_{n-1}}&=X^\mu F_{\mu\lambda_1\dots \lambda_{n-1}}
 \cr &\qquad +d\Lambda_{\lambda_1\dots\lambda_{n-1}}~.
\label{steady3}
\end{align}
The proof of the statement that all steady solitons are gradient solitons in section \ref{ssec:scaleconf} requires that $\bar S$ is invariant under diffeomorphisms and gauge transformations of $A$. Indeed, this is the case by the construction of $\bar S$.  This follows because the action $S$ is diffeomorphic invariant and the same applies for all the conditions needed to construct $\bar S$.  Moreover, all functionals used depend on $F$, which is sufficient condition to prove gauge invariance as well.

Returning to the case at hand, it is required to show that $\bar S_T$ is well-defined and has the required invariance properties. This has already been established for $ S$ and $\bar S$. It remains to focus on $S_{\mathrm{CS}}(A)$.
As we have assumed that $A\wedge^k F$ is globally defined, $S_{\mathrm{CS}}(A)$ is diffeomorphic invariant and so the same applies for the functionals $S_T$ and $\bar S_T$.  Moreover, $S_{\mathrm{CS}}(A)$ is invariant under the gauge transformations\footnote{ Note though that under  ``large'' gauge transformations
$A\rightarrow A'=A+\eta$
where $\eta$ is a closed but not exact $(n-1)$-form,  $S_{\mathrm{CS}}(A)$ may not be invariant.  Instead, it will transform as
$
S_{\mathrm{CS}}(A')=S_{\mathrm{CS}}(A)+\int_{M^D}\, \eta\wedge^k F$.
However, here for the arguments that follow, we do not require invariance under such transformations.}
\be
A\rightarrow A'=A+d\Lambda~,
\ee
provided that the gauge parameter $\Lambda$ is a globally defined  $(n-2)$-forms. This is a consequence of Stokes' theorem and of the closure of $F$, $d F=0$.  Thus, $S_T$ and $\bar S_T$ are invariant under such gauge transformations as well.

Having shown that $\bar S_T$ is diffeomorphic invariant and also invariant under gauge transformations of $A$ generated by exact forms, the argument in section  \ref{ssec:scaleconf}  can be repeated for $\bar S_T$ and for the steady M-flow solitons.  This establishes that all steady solitons (\ref{steady3}) are also gradient solitons and so they also  satisfy (\ref{grad3}).

Furthermore, set $c=-1/4$ and consider steady solitons  with $\Phi=0$. Also observe that in this case, upon setting   $X=a d\Phi$, the steady solitons (\ref{steady3}) with $\Phi=0$ satisfy the
gradient soliton equation (\ref{grad3}) upon setting $X=a d\Phi$.  In view of the result we have proven above that all steady solitons with $\Phi=0$ are gradient solitons, this will imply that either $X=a d\Phi$ or there exists a vector field $Y=X-ad\Phi$ that is Killing,  $\mathcal{L}_Y F=0$ and $\mathcal{L}_Y\Phi=0$.  The details are the same as those at the end of section \ref{ssec:scaleconf} and we shall not repeat them here.

\subsection{More general M-flows}

So far we have restricted our analysis to M-flows for which the coupling $b=4$.  This can be extended to M-flows with $b\not=4$ provided we introduce an axion field as in section \ref{ssec:axion}. The observation is that the analysis in the previous sections with the Chern-Simons term goes through provided we pick one $(n-1)$-form gauge potential, say $A^0$, to construct $S_{\mathrm{CS}}$ and at the same time add additional form field strengths of any degree. This means that we can replace the $c\,F^2$ term in the action (\ref{maction}) with
\be
c\, F^2\rightarrow c_0(F^0)^2+\sum_{k>0} c_k (F^k)^2~.
\ee
The analysis we have made all goes through without any additional alterations provided of course we introduce additional appropriate flows for the gauge potentials. In particular, we can base our analysis to the action (\ref{act2a}) and add a Chern-Simons term using the gauge potential $A^0$ of $F^0$. We can also modify $S_{\mathrm{CS}}$ by considering Chern-Simons style of terms that involve the form field strengths and gauge potentials of more than one fields.

\section{Conclusions}

In this paper, we provided a generalization of the Ricci-flow to gravity coupled to $k$-form potentials, the brane flows. The flow equations were inspired by the supergravity actions describing brane solutions. A particular case, which was the springboard of our analysis, is that of string theory 5-branes (NS5 and D5 branes), which couple magnetically to 2-form potential and describe the long range fields sourced by NS5 and D5-branes in 10 dimensions. Our brane flows holds more generally in any dimension and for general combination of form potentials. The supergravity actions typically contain Chern-Simons type terms and we showed that our analysis also generalizes to include such couplings. To the best of our knowledge, this is the first such generalization of the Ricci-flow. 

Following Perelman, we showed that the flow equations that preserve a specific dilaton-dependent volume enjoy monotonicity properties. Intriguingly, the volume that one keeps fixed is T-duality invariant, or related to it by S-duality. It would be interesting to understand the reason behind this and its implications. Is there a deeper role for string theory? 

We have also shown that all steady solitons are gradient solitons. In particular, steady solitons must admit a Killing vector that leaves invariant both the metric and the field strength. 
One may adjust the coupling in the flow equations such that the gradient scalar is actually the dilaton field. This means that the fixed point of the flow equation are solutions of the string theory effective action (when $d=10, p=4$). In the case of the fundamental string flows, this property was linked to the question of whether the fixed point of the two-dimensional sigma model is scale or conformal invariant and the monotonicity of the flow with the monotonicity of the worldsheet RG flow. It would be interesting to understand whether such interpretation is possible for the brane flows we discuss here. We note that the worldvolume theories of NS5 and D5 branes are not conformal when the type II supergravity equations hold (but note that in the M-theory limit, the worldvolume theory of the M5 branes is associated with the $(2,0)$ CFT). 

Our analysis was local and we have not explored the long time existence of the flows. It is well known that the Ricci flow can become singular at regions of positive curvature. For brane flows, the equations are modified by the fluxes and so the question has to be re-examined. Moreover, in the case of Ricci-flow  surgery the structure of Ricci solitons play a crucial role, at least in three dimensions,  in order to be able to continue the flow pass the singularities. It would be interesting to investigate this question in the context  brane flows. One generally expects that string theory resolves singularities and the fact that corresponding solitons may be related to branes suggests that string theory and brane physics will play crucial role in establishing the long time existence of the brane flows.

\vskip1cm
\noindent {\it {Acknowledgements:}} K.S. is supported in part by the STFC consolidated grant ST/X000583/1 ``New Frontiers in Particle Physics, Cosmology and Gravity."

\setcounter{section}{0}

 \appendix{}
 
 Here, we shall briefly describe  under which conditions the action $S_{\mathrm{CS}}(A)$ is well-defined.  We shall explore three possibilities that lead to the same conclusion. This is that $S_{\mathrm{CS}}(A)$ is well-defined provided that $\omega\equiv  A\wedge^k F$, $k\geq 1$,  must be globally defined on $M^D$ -- here well-defined means that $S_{\mathrm{CS}}(A)$ takes values in $\bR$ and its value for a given $A$ is independent from a wide range of choices in the structure of $M^D$ that have to be made to define the integral. This point will become increasingly clear below. We shall also point out that we can modify  $S_{\mathrm{CS}}(A)$ such that it becomes well-defined without $\omega$ be a globally defined form.  But such modifications change the field equations of $A$ and so they are not suitable for the purpose at hand.

To begin with the first approach,  let as consider a suitable open cover of $\{U_\alpha\}$ of $M^D$ such that $A$ patches as
\be
A_\alpha=A_\beta+d\chi_{\alpha\beta}~,
\ee
where $\chi$ is the patching gauge transformation defined at the intersection, $U_\alpha\cap U_\beta$, of the two open sets $U_\alpha$ and $U_\beta$. As a result, $\omega\equiv  A\wedge^k F$,  patches as
\be
\omega_\alpha=\omega_\beta+d\tau_{\alpha\beta}~,~~~\tau_{\alpha\beta}=-\tau_{\beta\alpha}~,
\label{gcs}
\ee
where $\tau_{\alpha\beta}$ is a $(D-1)$-form defined on $U_\alpha\cap U_\beta$.

Choosing a partition of unity $\{\rho_\alpha\}$ subordinate to $\{U_\alpha\}$ on $M^D$, the integral of $\omega$ can be defined as
\be
\int_{M^D} \omega\equiv \sum_\alpha \int_{U_\alpha} \rho_\alpha\, \omega_\alpha~,
\label{defi}
\ee
see {\it e.g.} \cite{BT} chapter II, where $\sum_\alpha \rho_\alpha=1$ at every point in $M^D$ and the sum over $\alpha$ is finite.
Although this integral is well defined for compact $M$ as each term in the sum is the integral of a function over a subset in Euclidean space and the sum over $\alpha$ is finite, it has to pass several other tests so it is defined unambiguously. For example, it can depend on the fact that  $\omega$ is defined locally and does not patch to a global form on $M^D$. It turns out that this is not the case. Indeed, multiply
(\ref{gcs}) with the product $\rho_\alpha \rho_\beta$ and sum over the two indices. This yields
\be
\sum_\alpha  \rho_\alpha\, \omega_\alpha=\sum_\beta \rho_\beta\, \omega_\beta+ \sum_{\alpha, \beta} \rho_\alpha\, \rho_\beta\, d \tau_{\alpha\beta}~,
\ee
where we have used the properties of a partition of unity, {\it i.e.}  $\sum_\alpha \rho_\alpha=1$ at every point in $M^D$.  But, the last term
in the above equation vanishes as $\tau_{\alpha\beta}=-\tau_{\beta\alpha}$.  Therefore, the integral of $\omega$ appears to be unambiguously defined.  However, this test is not enough. The value of the integral can depend on the choice of the partition of unity  used in (\ref{defi}).

To see this, consider another cover $\{V_{\alpha'}\}$ of $M$ and the partition of unity $\{\lambda_{\alpha'}\}$ subordinate to it. Then
\begin{align}
   \int^\rho_{M^D} \omega&\equiv \sum_\alpha \int_{U_\alpha} \rho_\alpha\, \omega_\alpha= \sum_\alpha \int_{U_\alpha} (\sum_{\beta'} \lambda_{\beta'})\,  \rho_\alpha\, \omega_\alpha= \sum_{\alpha, \beta'} \int_{U_\alpha\cap V_{\beta'}} \, \lambda_{\beta'} \rho_\alpha \,\omega_\alpha
   \cr
   &=\sum_{ \alpha, \beta'} \int_{ V_{\beta'}} \,  \rho_\alpha \,\lambda_{\beta'}\,(\omega_{\beta'}+d \psi_{\alpha\beta'})=\sum_{\beta'} \int_{V_{\beta'}} (\sum_\alpha \rho_\alpha) \, \lambda_{\beta'} \omega_{\beta'}+ \sum_{\alpha, \beta'} \int_{U_\alpha\cap V_{\beta'}} \, \lambda_{\beta'} \rho_\alpha \, d \psi_{\alpha\beta'}
   \cr
   &= \int^\lambda_{M^D} \omega+\sum_{\alpha, \beta'} \int_{U_\alpha\cap V_{\beta'}} \, \lambda_{\beta'} \rho_\alpha \, d \psi_{\alpha\beta'}~,
   \label{depunit}
\end{align}
where we have used that $\lambda_{\beta'} \rho_\alpha \,\omega_\alpha$ has support in $U_\alpha\cap V_{\beta'}$ and so in $V_{\beta'}$.  The last term in the above expression can be written in several ways but generically it does not vanish.  The only way that this term vanishes is if $\omega$ can be ``improved'' to a globally defined form on $M$. This requires that the associated Cech cohomology cycle becomes trivial. A sufficient condition for this is that
$\tau$ in (\ref{gcs})  satisfies the trivial  cocycle condition
\be
\tau_{\alpha\beta}+\tau_{\beta\gamma}+\tau_{\gamma\alpha}=0~.
\label{2cycle}
\ee
(As we shall explain below, it is always possible to arrange such that this is a necessary condition as well.)
In this case, $\omega$ can be ``improved'' using a gauge transformation at every patch to define a globally defined form $\tilde\omega$ as
\be
\tilde \omega_\alpha=\omega_\alpha-d \sum_\alpha \rho_\gamma \tau_{\alpha\gamma}~.
\label{0cycle}
\ee
Indeed,
\be
\tilde\omega_\alpha-\tilde \omega_\beta=\omega_\alpha-\omega_\beta-d \sum_\gamma \rho_\gamma (\tau_{\alpha\gamma}-\tau_{\beta\gamma})=d\tau_{\alpha\beta}- d(\sum_\gamma \rho_\gamma \tau_{\alpha\beta})=0~.
\label{1cycle}
\ee
So $\tilde\omega$ is globally defined and it can be used to define the integral.  In such a case, the contribution
of the terms that contain $\psi$ in the calculation presented in (\ref{depunit}) vanish and the integral is independent from the choice of partition of unity utilized  to define it.

To explain why the cocycle condition (\ref{2cycle}) can be arranged to be necessary as well, suppose that $\omega$ is a 3-form -- the proof can be adapted to forms of any degree but to avoid introducing additional machinery let us focus on 3-forms. In general the patching conditions  at the intersection of two or more open sets $U_\alpha$ are
\begin{align}
&\omega_{\alpha_0}-\omega_{\alpha_1}=d\tau_{\alpha_0\alpha_1}~,~~\text{on}~~U_{\alpha_0\alpha_1}\equiv U_{\alpha_0}\cap U_{\alpha_1}~,
\cr
&\tau_{\alpha_0\alpha_1}+\tau_{\alpha_2\alpha_0}+\tau_{\alpha_1\alpha_2}=d\chi_{\alpha_0\alpha_1\alpha_2}~,~~\text{on}~~U_{\alpha_0\alpha_1\alpha_2}~,
\cr
&\chi_{\alpha_0\alpha_1\alpha_2}-\chi_{\alpha_3\alpha_0\alpha_1}+\chi_{\alpha_2\alpha_3\alpha_0}-\chi_{\alpha_1\alpha_2\alpha_3}
=d\psi_{\alpha_0\alpha_1\alpha_2\alpha_3}~,~~\text{on}~~U_{\alpha_0\alpha_1\alpha_2\alpha_3}~,
\cr
&\psi_{\alpha_0\alpha_1\alpha_2\alpha_3}+\psi_{\alpha_4\alpha_0\alpha_1\alpha_2}+\dots =c_{\alpha_0\alpha_1\alpha_2\alpha_3\alpha_4}~,~~\text{on}~~U_{\alpha_0\alpha_1\alpha_2\alpha_3\alpha_4}~,
\label{cocyclon}
\end{align}
where $\tau$, $\chi$ and $\psi$ are forms of decreasing degree and $c$ are constants. The forms $\tau$, $\chi$ and $\psi$ are not uniquely defined -- they are defined up to an exact form at the corresponding intersection. It is clear that we have used the Poincar\'e lemma at every stage to  construct the sequence of cocycle conditions in (\ref{cocyclon}).  For example, from the first patching condition, we deduce that
\be
d(\tau_{\alpha_0\alpha_1}+\tau_{\alpha_2\alpha_0}+\tau_{\alpha_1\alpha_2})=0~.
\ee
and Poincar\'e lemma is used to solve this condition to  establish the second cocycle condition in (\ref{cocyclon}) and so on. One may worry that the form in the parenthesis in the above equation is closed but not exact and therefore the second cocycle condition in (\ref{cocyclon}) does not follow from the above equation. However, this is not the case as there are always covers on $M$, the good covers, that the Poincar\'e lemma can be used at all open sets and their finite intersections, see \cite{BT}. Therefore, these calculations are performed on good covers or on a cover that for the specific form $\omega$ one can use the Poincar\'e lemma at every stage.

 The claim is that if any of the right hand side of the cocycle conditions in (\ref{cocyclon}) vanishes, then $\omega$ can be ``improved'' to a globally defined form on $M$.  In particular, if any of the right hand side of the cocyle conditions below that for $\tau$ vanishes, then $\tau$ can be ``improved'' to satisfy (\ref{2cycle}).
We start by considering the case where all constants $c$ vanish, $c=0$.  Possible extension where $c \neq 0$ are discussed afterwards. In the case at hand, observe that the ``improved'' $\chi$ given by
\be
\tilde \chi_{\alpha_0\alpha_1\alpha_2}=\chi_{\alpha_0\alpha_1\alpha_2}+ d\sum_{\gamma} \rho_{\gamma} \psi_{\alpha_0\alpha_1\alpha_2\gamma}~,
\ee
satisfies the trivial cocycle condition 
\be
\tilde \chi_{\alpha_0\alpha_1\alpha_2}-\tilde \chi_{\alpha_3\alpha_0\alpha_1}+\tilde \chi_{\alpha_2\alpha_3\alpha_0}-\tilde\chi_{\alpha_1\alpha_2\alpha_3}=0~.
\label{3cycle}
\ee
As the improvement of $\chi$ is the exterior derivative of a form, the cocycle condition for $\tau$ in (\ref{cocyclon}) can now be rewritten as
\be
\tau_{\alpha_0\alpha_1}+\tau_{\alpha_2\alpha_0}+\tau_{\alpha_1\alpha_2}=d\chi_{\alpha_0\alpha_1\alpha_2}=d\tilde\chi_{\alpha_0\alpha_1\alpha_2}~,~~~
\text{on}~~U_{\alpha_0\alpha_1\alpha_2}~.
\ee
This allows to ``improve'' $\tau$ by setting
\be
\tilde \tau_{\alpha_0\alpha_1}=\tau_{\alpha_0\alpha_1}-d \sum_{\gamma} \rho_{\gamma} \tilde \chi_{\alpha_0\alpha_1\gamma}~.
\ee
It can be shown using (\ref{3cycle})  that $\tilde \tau$ satisfies the trivial cocycle condition (\ref{2cycle}) which brings us back to the computation in 
(\ref{1cycle}) that leads to an ``improved'' $\omega$ that it is globally defined on $M$.

A second approach to define $S_{\mathrm{CS}}(A)$ has been explored sometime ago by O. Alvarez \cite{OA} -- for a more recent detailed account see \cite{FT}. See also \cite{Colombo:2025yqy} for a discussion in the context of the contribution of Chern-Simons type terms in the on-shell supergravity action.   We shall not give details of this approach but the definition of $\int_{M^D} \omega$  involves the use of the patching conditions of $\omega$ in increasingly higher intersections of open sets to define the integral -- notice that in (\ref{defi}) we have not considered contributions from the intersections of open sets. At the end, this approach succeeds to prove that $\exp (i S_{\mathrm{CS}}(A))$ is well defined for a suitably defined $S_{\mathrm{CS}}(A)$, i.e. independent of choices,  provided that $\omega$ represents a (non-trivial) integral class of $M^D$. In other words, $S_{\mathrm{CS}}(A)$ is not well-defined. Instead it is defined up to a possible ambiguity that involves the addition of an integer. This is not good enough for our purpose as we rely on the monotonicity of $S_{\mathrm{CS}}(A)$ to prove our results and a shift in the value of $S_{\mathrm{CS}}(A)$, which depends on the details of evaluating the integral,  can uncontrollably  increase or decrease its value.

A third approach to the problem is to use the {\it collating formula}, proposition 9.5 page 102 in \cite{BT}, which associates to every $\breve {\text C}$ech-de-Rham  $n$-cochain $a=\sum_{i=0}^n a_i$ a globally defined form $f(a)$ on $M^D$, where $\{a_i; i=0,\dots, n\}$ is a collection of forms of degree $n-i$ defined on a $i$-th intersection $U_{\alpha_0 \alpha_1\dots \alpha_i}$ of open sets. To simplify the exposition, consider the case that the gauge potential $A$ is a 1-form with patching conditions
\begin{align}
&A_{\alpha_0}-A_{\alpha_1}=d \lambda_{\alpha_0\alpha_1}~,~~~\text{on}~~~ U_{\alpha_0\alpha_1}~,~~~\lambda_{\alpha_0\alpha_1}+\lambda_{\alpha_2\alpha_0}+\lambda_{\alpha_1\alpha_2}=c_{\alpha_0\alpha_1\alpha_2}~,~~~
\text{on}~~~U_{\alpha_0\alpha_1\alpha_2}~,
\cr
&c_{\alpha_0\alpha_1\alpha_2}-c_{\alpha_3\alpha_0\alpha_1}+c_{\alpha_2\alpha_3\alpha_0}-c_{\alpha_1\alpha_2\alpha_3}
=0~~~\text{on}~~~U_{\alpha_0\alpha_1\alpha_2\alpha_3}~,
\end{align}
i.e. the $\breve {\text C}$ech-de-Rham 1-cochain is $A=\{A, \lambda, c\}$. An application of the collating formula reveals that
\be
f(A)_\alpha=A_\alpha-\sum_\beta \rho_\beta\, d\lambda_{\alpha\beta}-\sum_\beta d\rho_\beta\, \lambda_{\alpha\beta}-\sum_{\beta\gamma} \rho_\beta\, d\rho_\gamma\, c_{\alpha\beta\gamma}~,
\ee
is a globally defined form on $M^D$, i.e. $f(A)_\alpha=f(A)_\beta$ on $U_{\alpha\beta}$. Therefore, it is tempting to define the Chern-Simons style of term as
\be
\tilde S_{\mathrm{CS}}(A)\equiv  \sum_\alpha \int_{U_\alpha} \rho_\alpha\, f(A)_\alpha \wedge^k F(A)_\alpha~.
\label{defi2}
\ee
Notice that $\tilde S_{\mathrm{CS}}(A)$ is well-defined as $f(A)\wedge^k F$ is a globally defined form on $M^D$ and in this definition contributions from multiple intersections of open sets are included. However, a calculation of the exterior derivative of $f(A)$ reveals that
\be
df(A)_\alpha=F_\alpha-\sum_{\beta\gamma} d\rho_\beta\wedge d\rho_\gamma\, c_{\alpha\beta\gamma}~.
\ee
As $\sum_{\beta\gamma} d\rho_\beta\wedge d\rho_\gamma\, c_{\alpha\beta\gamma}$ is the $\breve {\text C}$ech representative of the de-Rham class of $F_\alpha$, see \cite{BT}, their difference is expected to be an exact form that in this case is given by the exterior derivative of $f(A)$. This is equivalent to introducing a source term $A_0$, such that the  de-Rham class of $F$ can be represented by either $dA$ or $dA_0$, and to defining 
\be
\tilde S_{\mathrm{CS}}(A)\equiv  \sum_\alpha \int_{U_\alpha} \rho_\alpha\, (A-A_0)_\alpha \wedge^k F(A)_\alpha~.
\label{defi3}
\ee
Of course, this gives a well defined expression for $\tilde S_{\mathrm{CS}}(A)$ but the addition of the source term will modify the field equations for $A$ that we have used to carry out our analysis. Although for simplicity, the argument above has been carried out for $\breve {\text C}$ech-de-Rham 1-cochains, it can be extended to the more general case of  $n$-cochains.

\bibliographystyle{unsrt}

\end{document}